**Exploring the Impacts of Air Quality on Travel Behavior and Activity Participation: Evidence from Travel Diary Surveys in Northern Utah**

**Fariba Soltani Mandolakani**
Department of Civil and Environmental Engineering
Utah State University, Logan, UT 84322-4110
Email: fariba.soltani@usu.edu
ORCID: 0009-0005-9089-0514

**Mahyar Vahedi Saheli**
Department of Civil and Environmental Engineering
Utah State University, Logan, UT 84322-4110
Email: mahyar.vahedi@usu.edu
ORCID: 0000-0002-8267-8354

**Patrick A. Singleton** (corresponding author)
Department of Civil and Environmental Engineering
Utah State University, Logan, UT, 84322-4110
Email: patrick.singleton@usu.edu
ORCID: 0000-0002-9319-2333

**ABSTRACT**

In this study, we explored whether and how area-wide air pollution affected individuals' activity participation and travel behaviors, and how these effects differed by neighborhood context. Using multi-day travel survey data provided by 403 adults from 230 households in a small urban area in northern Utah, US, we analyzed a series of 20 activity and travel outcomes. We investigated the associations of three different metrics of (measured and perceived) air quality with these outcomes, separately for residents of urban and suburban/rural neighborhoods, and controlled for personal and household characteristics. Our models found some measurable changes in activity and travel patterns on days with poor air quality. In urban areas, people engaged in more mandatory (work/school) activities, whereas there was no discernible change in suburban/rural areas. The total travel time for urban residents increased, driven by increases in trip-making and travel time by public modes (bus) and increases in travel time by private modes (car). On the other hand, suburban/rural residents traveled shorter total distances (mostly through lower vehicle miles traveled), and there was a notable uptick in the probability of being an active mode user (walk/bike). Air quality perceptions also seemed to play a role, at least for urban residents who walked/biked longer distances, rode the bus for longer distances/times, and drove fewer miles on days with worse perceived air pollution. Overall, the results are somewhat encouraging, finding more evidence of altruistic than risk-averse travel behavioral responses to episodes of area-wide air pollution; although, more research is needed.

**Keywords**: Travel behavior, Activity choices, Air quality, Air pollution, Travel surveys.

# INTRODUCTION

Despite plentiful knowledge about the effects of transportation on air quality (*1*), research has rarely investigated the reverse link: How does air pollution or air quality perceptions affect individuals' travel behaviors? Such insights would be useful for evaluating and designing air quality improvement policy measures, including those that attempt to reduce polluting automobile use—and promote the use of active and sustainable modes (walking, bicycling, and public transit)—through "hard" and "soft" policies (*2*). Many policies are assumed to operate on and influence individuals and their transportation choices. Thus, knowledge of the effects of air pollution (and perceptions thereof) on individual-level travel behaviors is important. Furthermore, there are complex behavioral motivations at play during episodes of poor air quality: altruism (driving less and riding transit more to avoid contributing to air pollution) versus risk-aversion (walking and riding transit less to avoid exposure to air pollution) (*3*). Studying travel behavioral sensitivities to air pollution advances understanding of decision-making under risk.

A limited but growing literature studies the effects of regional air quality levels on travel behaviors. Focusing just on research in the US, findings are somewhat inconsistent and location-specific. While some studies find no significant change (or a modest decline) in motor vehicle traffic volumes on days with air quality alerts or elevated levels of air pollution, other research suggests that driving may increase on such days (*4-6*). Ozone pollution alerts increased public transit usage in San Francisco (*5*) but not in Chicago (*7*). One fairly consistent finding (across four studies) is that high levels of air pollution tends to decrease active transportation, as measured by bicycle, pedestrian, and non-motorized trail counts (*6,8-10*). However, all of the above-mentioned studies used secondary sources (traffic counts) and aggregate analyses of traffic volumes. These methods can only suggest (but not explain) why and how travel behaviors are affected by area-wide air pollution (if at all).

Instead, measuring individual-level travel behaviors could be more informative. Some limited travel survey-based research has been done in the US. Of two such studies in Atlanta, one found decreases in miles driven but not trips taken on ozone alert days (*11*), while the other found that smog alerts did not significantly decrease household vehicle miles traveled (*3*). Overall, most prior research has focused on summer ozone levels (rather than wintertime particulate matter) in a few large cities. Studying individual responses can also help control for some other personal and locational factors that contribute to heterogenous travel and activity behaviors. In particular, we anticipate that the influence of air quality on activity participation and travel behavior may differ across built environment contexts (e.g., urban and suburban/rural neighborhood types), as such areas have different transportation options and accessibilities to destinations that may facilitate or constrain behavioral responses to air pollution.

Our study's primary objective is to determine whether and how measured (or perceived) area-wide air pollution affects individuals' daily travel behaviors. A secondary objective is to assess how these associations differ by neighborhood type (urban vs. suburban/rural). To achieve these goals, we analyzed a series of activity participation and travel behavior outcomes taken from a multi-day travel diary survey (on winter days of varying air quality) in a small urbanized area in northern Utah troubled by periodic high concentrations of PM2.5. In the following sections, we summarize our data and methods, and then discuss our results and interpret key findings.

## DATA AND METHODS

**Study Area**

Our study area is Cache Valley, a region in northern Utah characterized by its distinctive geography, situated at a high elevation between two mountain ranges. This unique topography creates ideal circumstances for wintertime temperature inversions, leading to a significant accumulation of particulate matter and other air pollutants in the lower atmosphere. Also, at the time of the study, Cache Valley was designated as a non-attainment area for PM2.5 (this status was removed in 2021). The region regularly experiences air pollution in winter, and its air quality is sometimes the worst in the state of Utah and even in the entire nation (*12*). Residents of Cache Valley often expect wintertime air pollution, and air quality alerts (*13*) and related travel demand management messages (*14*) are regularly distributed through local news media. Consequently, Cache Valley is an excellent location for studying the connections between travel behavior and air pollution, because of how frequently elevated air pollution levels occur and the moderate awareness of this issue among the local population.

**Data Collection**

During the winter of 2019 (January–March), we conducted an online panel travel diary survey targeting households in Cache Valley. To ensure participants were recruited from a diverse range of built environment contexts, we first classified US Census block groups into three strata—very urban, somewhat urban, and suburban/rural—based on their scores on four variables (housing unit density, intersection density, job access by automobile, and transit frequency) taken from the Smart Location Database version 2.0 (*15*). Next, we used stratified random sampling to select block groups to fulfill our quotas of 2,000 households in each of the very and somewhat urban groups, and 4,000 households in the suburban/rural group. Finally, we obtained residential addresses for the selected block groups, and mailed each housing unit a paper letter containing a description of the study and a website link to sign-up every adult member of the household.

The data collection process was organized into three distinct phases.

1. **The initial survey:** Once participants enrolled in the study, they were asked to answer a set of questions regarding household composition, demographics, and transportation-related information.
2. **Travel diary surveys:** In the second phase, participants were required to complete three rounds of two-day travel diary surveys. We strategically scheduled these rounds over the course of several weeks to attempt to encompass a range of (good, moderate, and unhealthy) air quality conditions, using day-ahead air quality forecasts. (In this way, we tried to take advantage of a natural experiment.) During this phase, each participant recorded detailed information about every trip undertaken on the survey day, including departure and arrival times, modes of transportation, locations, and trip purposes.
3. **Final survey:** Following the completion of the travel diary surveys, each participant was asked to participate in a final survey. This survey asked questions about various psycho-social factors, such as attitudes, values, and norms related to transportation choice and air quality. We did not use the responses from the final survey in this paper's analyses.

In total, an invitation was sent to 8,376 households. From this, 255 households (consisting of 479 adults) completed the initial survey, a response rate of 3%. In the end, 189 households (337 adults) completed the final survey, for a 25–30% attrition rate. The analyses presented here contain responses by 403 adults from 230 households, including anyone who completed at least one travel diary survey. For more details on the data collection effort, see Humagain and Singleton (*16*).

*Dependent Variables*

The dependent variables (DVs) in this study were measures of activity participation and travel behavior derived from the self-reported online travel diaries. We performed a significant amount of data cleaning on the survey responses, removing duplicate and incomplete entries, geocoding places, and calculating travel times and distances traveled using several Google Maps APIs. From the cleaned travel diary data, we constructed daily totals of each individual's activity participation—the number of out-of-home activities, by activity category (mandatory, discretionary, or semi-mandatory/discretionary)—and travel behaviors: the number of trips made, distance traveled, and travel time, all segmented by mode category (active, public, private). These categories are defined in Table 1.

At the end of this process, we realized that many of our DVs had a preponderance of zeros, due to either not traveling on the survey day or not using certain modes. Therefore, we constructed a series of sequential DVs, where earlier activity/travel decisions split the data, and models of later outcomes only used a subset of the data. The first binary DV was whether or not the respondent stayed at home (did not travel). Next, if false (did travel), a series of DVs represented daily activity participation and all-mode travel outcomes. Then, three binary DVs assessed whether or not the respondent used each mode category. Finally, if true, the three remaining travel outcome DVs (trips, distance traveled, travel time) were calculated for people who did use each mode. Table 1 presents sample sizes and descriptive statistics for each of the study's 20 DVs.

*Independent Variables*

Given this study's focus on air pollution, we used several different air quality metrics as independent variables (IVs). Measured air quality was assessed using the Air Quality Index (AQI), a 0–500 measure of air pollution concentrations (*17*). To examine potential non-linear effects, we also categorized AQI based on the well-publicized color: green ("good" AQI = 0–50), yellow ("moderate" AQI = 51–100), and orange ("unhealthy" AQI = 101+). Despite our best attempt to capture a range of air quality conditions, most observations occurred on days with green or yellow air. Perceived air quality was measured by a response at the end of each travel diary survey, where respondents rated the air quality on a 1–5 scale (1 = great, good, fair, bad, 5 = terrible). AQI and perceived air quality were positively but not perfectly correlated (0.30). Together, these three air quality IVs (AQI number, AQI category, perceived air quality) were used to investigate variations in the relationships with the activity and travel behavior DVs.

Although not the primary focus of this study, we also considered as IVs other control variables pertaining to respondents' personal and household characteristics. Personal characteristics included self-reported age, race/ethnicity, gender, educational attainment, and student and worker statuses. Household characteristics included housing type, household income, household composition (children, adults), and mobility tools (bicycles, motor vehicles).

Additionally, we included home neighborhood type as a binary measure of the built environment; this was based on our block group sampling strategy (very/somewhat urban vs. suburban/rural) discussed earlier. Table 2 shows descriptive statistics for the IVs. Figure 1 maps the neighborhood type of the Census block groups that contained the home locations of study participants.

**Table 1: Descriptive statistics of the dependent variables**

|  |  | *Categorical* |  | *Continuous* |  |
|---|---|---|---|---|---|
| *Dependent variable* | N | # | % | Mean | SD |
| Stayed at home (did not travel) | 2,044 |  |  |  |  |
|     True |  | 220 | 10.76 |  |  |
|     False |  | 1,824 | 89.24 |  |  |
| Activity participation (#) | 1,824 |  |  |  |  |
|     Total out-of-home |  |  |  | 2.56 | 1.74 |
|     Mandatory[a] |  |  |  | 1.01 | 0.90 |
|     Semi-mandatory/discretionary[b] |  |  |  | 0.74 | 1.20 |
|     Discretionary[c] |  |  |  | 0.82 | 1.14 |
| Travel outcomes, all modes | 1,824 |  |  |  |  |
|     Number of trips (#) |  |  |  | 4.32 | 2.43 |
|     Distance traveled (miles) |  |  |  | 25.23 | 47.94 |
|     Travel time (minutes) |  |  |  | 65.73 | 56.84 |
| Used mode on travel day | 1,824 |  |  |  |  |
|     Active modes: True |  | 295 | 16.17 |  |  |
|       False |  | 1,529 | 83.83 |  |  |
|     Public modes: True |  | 151 | 8.28 |  |  |
|       False |  | 1,673 | 91.72 |  |  |
|     Private modes: True |  | 1,703 | 93.37 |  |  |
|       False |  | 121 | 6.63 |  |  |
| Active mode[d] users | 295 |  |  |  |  |
|     Number of trips (#) |  |  |  | 2.31 | 1.32 |
|     Distance traveled (miles) |  |  |  | 4.66 | 20.33 |
|     Travel time (minutes) |  |  |  | 34.83 | 28.89 |
| Public mode[e] users | 151 |  |  |  |  |
|     Number of trips (#) |  |  |  | 1.65 | 0.69 |
|     Distance traveled (miles) |  |  |  | 5.70 | 11.56 |
|     Travel time (minutes) |  |  |  | 23.64 | 21.82 |
| Private mode[f] users | 1,703 |  |  |  |  |
|     Number of trips (#) |  |  |  | 4.08 | 2.39 |
|     Distance traveled (miles) |  |  |  | 25.71 | 48.48 |
|     Travel time (minutes) |  |  |  | 62.27 | 57.78 |

[a] Mandatory activities include: work, school, work- or school-related
[b] Semi- activities include: civic or religious, drop off or pick up passenger, other errands or appointments, service private vehicle
[c] Discretionary activities include: eat meal at restaurant, social or entertainment, outdoor or indoor exercise, shopping
[d] Active modes include: walk, bicycle
[e] Public modes include: school bus, local bus
[f] Private modes include: car/van/truck/SUV driver or passenger, motorcycle/scooter/moped

**Table 2: Descriptive statistics of the independent variables**

| Independent variable | Categorical | | Continuous | |
|---|---:|---:|---:|---:|
| | # | % | Mean | SD |
| **Household characteristics** | | | | |
| Housing type: Single-family | 304 | 75.62 | | |
|    Multi-family | 98 | 24.38 | | |
| Household income: < $35,000 | 96 | 24.00 | | |
|    $35,000 to $74,999 | 158 | 39.50 | | |
|    ≥ $75,000 | 122 | 30.50 | | |
|    Unknown | 24 | 6.00 | | |
| Number of children | | | 0.98 | 1.36 |
| Number of adults | | | 2.02 | 0.64 |
| Number of bicycles | | | 2.08 | 1.91 |
| Number of motor vehicles | | | 1.96 | 0.90 |
| Neighborhood type[a]: Urban | 237 | 58.81 | | |
|    Suburban or rural | 166 | 41.19 | | |
| **Personal characteristics** | | | | |
| Age: 18 to 34 years | 182 | 45.61 | | |
|    35 to 54 years | 132 | 33.08 | | |
|    ≥ 55 years | 85 | 21.30 | | |
| Race/ethnicity: White-alone | 368 | 92.93 | | |
|    Non-white or multiple | 28 | 7.07 | | |
| Gender: Male | 190 | 47.50 | | |
|    Female | 210 | 52.50 | | |
| Education: Less than bachelor | 157 | 39.15 | | |
|    Bachelor's degree or higher | 244 | 60.85 | | |
| Student: No | 313 | 78.05 | | |
|    Yes | 88 | 21.95 | | |
| Worker: Yes | 304 | 75.81 | | |
|    No | 97 | 24.19 | | |
| **Air quality measures** | | | | |
| Air quality index (AQI) | | | 47.77 | 21.14 |
|    Green (0 – 50) | 1,008 | 49.32 | | |
|    Yellow (51 – 100) | 1,013 | 49.56 | | |
|    Orange (101 – 150) | 23 | 1.13 | | |
| Perceived air quality[b] | | | 2.51 | 0.94 |

[a] Classification of block groups based on housing unit density, intersection density, job access by automobile, transit frequency
[b] Rating of air quality, 1 = great, 5 = terrible

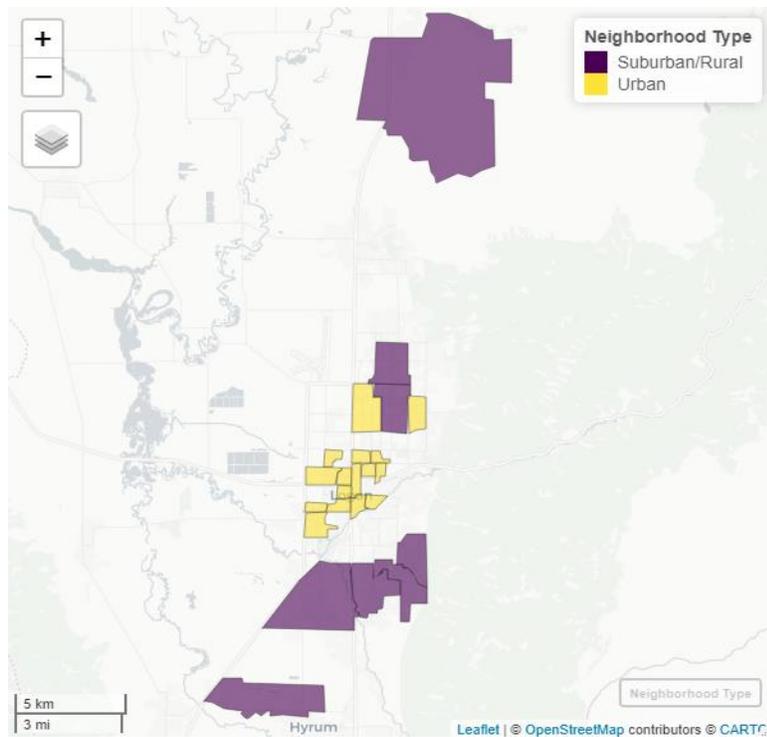
**Figure 1: Map of sampled Census block groups by neighborhood type**

**Analysis Methods**

As described earlier, we used 20 different DVs representing different activity and travel outcomes. Three different types of statistical models were applied, depending on the type of DV.

- For each of the continuous DVs (distance traveled, travel time), we used a log-linear regression model. In this model, the original DV is transformed using the natural log, which we found to better fit our data and yield a more normal distribution. We also added 1 to the travel outcomes before taking the natural log, to avoid issues where ln 0 is undefined, and to avoid negative outputs.
- For each of the binary DVs (stayed at home, used each mode), we applied logistic regression, also known as the binary logit model.
- For each of the count DVs (number of activities, number of trips), we started by considering the Poisson regression model, a common choice for modeling non-negative integer values. However, the Poisson model assumes that the variance is equal to the mean, which is not always realistic. Instead, one can allow for over-dispersion (variance > mean) by adding an extra parameter to the variance equation that is either a linear or quadratic function of the mean, resulting in the quasi-Poisson or negative binomial models, respectively. We tried all three options and found that the quasi-Poisson models had better fits to the data, so we used quasi-Poisson regression for all count DVs.

Finally, we must mention that we actually estimated three sets of models, one set for each of the ways of representing air quality (AQI number, AQI category, perceived air quality). Also, to be clear, we interacted the air quality variables with our neighborhood type variable. Doing this allowed us to investigate how different types of neighborhoods (urban vs. suburban/rural) influence the manner in which air quality affects travel behavior changes.

**RESULTS AND DISCUSSION**

**Overall Results**
Table 3 presents abbreviated model results—only the signs of statistically significant coefficients—for the 20 models (one for each of the activity participation and travel behavior DVs) containing the AQI representation of air quality. We also inspected model results for those using AQI category and perceived air quality; the abbreviated results are virtually identical for non-air quality IVs. More detailed results for the various air quality measure are contained in the next section. Here, we briefly report some key findings for the other household and personal characteristics IVs, since they are not core to the study's objective. (Full model results are available from the authors upon request.)

Compared to people living in single-family detached houses, people living in multi-family housing participated fewer mandatory activities, but more semi-mandatory/discretionary and discretionary activities (and more total activities). They also tend to make more private mode (and total all) trips, and have higher distance traveled for public modes. On the other hand, the number of trips by active modes and the odds of using public modes were lower for residents of multi-family housing.

Income level also played a significant role in travel behaviors (but not activity participation). Respondents in lower-income (< $35,000) households were more likely to stay at home. If they did travel, they were more likely to use private modes (automobile driver or passenger) as their transportation means. They also made a higher number of trips with public modes. Meanwhile, members of lower-income households tended to travel less in both time and distance (overall, and for private modes). Lower-income active mode users made fewer trips and traveled shorter distances and for less time, while lower-income public mode users actually made more trips. In comparison, there were fewer associations for people in high-income households (≥ $75,000). These individuals had higher total distance traveled by all modes. Also, members of high-income households took fewer trips by active and public modes, if they were users of these modal categories.

Household composition also affected activity participation and travel behaviors. People in households with more children participated in more total activities—especially semi-mandatory/discretionary—but fewer mandatory activities. Compared to people with fewer children, these individuals made more trips and had higher distance traveled and travel time across all modes. Also, they tended to use private mode more and active and public modes less. However, among respective mode users, people in households with more children had higher distance traveled by active modes, higher travel distance and time by public modes, and more trips/distance/time using private modes. In contrast, having more adult members of the household was associated with a greater chance of staying home. The total number of activities (including mandatory and discretionary activities) tended to be less for this group. Besides, total distance traveled and travel time was also less for them. Regarding private modes, the number of trips, distance traveled, and travel time tended to be less for people in households with more adults. Interestingly, the odds of using public modes increased with the number of adults.

# Table 3: Abbreviated model results for models with AQI

| Variable | SH | Activities T | M | S | D | All modes NT | DT | TT | Active modes U | NT | DT | TT | Public modes U | NT | DT | TT | Private modes U | NT | DT | TT |
|---|---|---|---|---|---|---|---|---|---|---|---|---|---|---|---|---|---|---|---|---|
| Housing type: Multi-family |  | + | − | + | + | + |  |  |  | − |  |  |  |  | + |  |  | + |  |  |
| Household income: < $35,000 | + |  |  |  |  | − | − |  | − | − | − |  | + |  |  |  | + |  | − | − |
| ≥ $75,000 |  |  |  |  |  | + |  |  |  | − |  |  | − |  |  |  |  |  |  |  |
| Unknown |  | + |  |  |  | + | + |  |  |  | − |  |  |  |  |  |  | + | + |  |
| Number of children |  | + | − | + |  | + | + | + | − |  | + |  | − |  | + | + | + | + | + | + |
| Number of adults | + | − | − |  | − | − | − |  |  |  |  |  | + |  |  |  |  | − | − | − |
| Number of bicycles | − |  |  |  |  | + |  |  | + | + |  | + | + |  |  |  | − |  |  |  |
| Number of motor vehicles | − |  | + |  |  | + |  |  |  | − | − | − |  |  |  |  | + |  | + | + |
| Age: 35 to 54 years |  | + | − | + |  | + |  |  |  | − | − | − | − |  | − |  | + | + |  |  |
| ≥ 55 years |  | + | − | + | + | + |  |  |  | − |  |  | − | − | − |  | + |  |  | + |
| Race/ethnicity: Non-white or multiple |  |  | − | + |  |  |  |  | − | − | − |  | − |  |  | + |  |  |  |  |
| Gender: Female |  | + | − | + | + | + | − |  | + |  |  |  | − |  | − | − | + | − | − |  |
| Education: Less than bachelor | + | − |  | − |  | − | − | − | − | + |  |  |  | − |  |  |  | − | − |  |
| Student: Yes | − | + | + | − | − | + |  |  | + | + | − | + | + |  |  |  | − |  |  |  |
| Worker: No | + | + | − | + | + | + |  |  | − | − | − | − | − |  |  |  | + |  |  |  |
| Neighborhood type: Suburban or rural |  |  |  |  |  | + | + | − |  | − | − | − | − |  |  |  | + |  | + | + |
| AQI: Urban |  |  | + |  |  |  | + |  |  |  |  |  | + |  | + |  |  |  | + | + |
| AQI: Suburban or rural |  |  |  |  |  |  | − |  | + |  |  | + |  |  |  |  |  |  | − |  |

Statistical significance: + $p < 0.10$ and $B > 0$, − if $p < 0.10$ and $B < 0$; blank if $p > 0.10$.

SH = stay at home; T = total, M = mandatory, S = semi-mandatory/discretionary, D = discretionary

U = user, NT = number of trips (#), DT = distance traveled (miles), TT = travel time (minutes)

The holding of more mobility tools like bicycles or motor vehicles was linked to some activity and travel outcomes. People with access to more bicycles and more motor vehicles were less likely to stay at home. Individuals with access to bicycles had a higher total number of trips and total travel time, and individuals with access to motor vehicle had higher total distance traveled. Those in households with more bicycles were more likely to use active and public modes, and less likely to use private modes. They also traveled longer distances using active modes. In contrast, individuals with access to more motor vehicles traveled longer distances overall, were less likely to use active or private modes, and were more likely to use private modes. Also, the distance traveled and travel time via private modes were higher. Regarding activities, individuals with access to motor vehicles participated in more mandatory activities.

Regarding age effects on activity participation and travel behaviors, compared to younger adults (below 35), middle-aged and older adults participated more in total activities, especially more semi-mandatory/discretionary, and fewer mandatory activities. Additionally, individuals older than 55 years participated in more discretionary activities. Both groups had a higher total number of trips. Increased age seemed to be correlated with reduced trip-making by active modes, a decreased odds of using public modes, and shorter distance traveled by public modes. Additional age-related results include fewer trips by public modes and longer travel time by private mode for adults older than 55 years, as well as shorter distances traveled and travel times by active mode.

Differences were observed in activity/travel behaviors based on self-identified race/ethnicity. People selecting one-or-more non-white racial/ethnic categories tended to participate in fewer mandatory and more semi-mandatory/discretionary activities, spent less time traveling in total, and used active and public modes less. Also, the number of trips by active mode decreased for these individuals, and they spent more time traveling using public modes.

Gender also had an impact on activity participation and travel behavior. People identifying as female did fewer mandatory activities and more semi-mandatory(discretionary) and discretionary activities (and total activities, overall). While women made more total trips, those trips tended to be shorter (shorter total distance traveled); the same trend was true for women automobile users with shorter travel distances and times. Women were more likely to use active modes and less likely to use public modes. The distance traveled and time by public mode were also less for women than for men.

Some effects were found for educational attainment. Respondents without a bachelor's degree were more likely to stay at home, and those who traveled made fewer total and semi-mandatory/discretionary trips. They also spent less time traveling, traveled shorter distances, and made fewer trips overall. Meanwhile, they were less likely to use active transportation modes, but if they used it, they made more trips by active modes. These individuals traveled shorter distances with public modes, and had fewer trips and shorter distance traveled with private modes.

Student and worker status indicators were also connected to activity and travel outcomes. Students were less likely to stay at home and tended to do more mandatory and total activities but fewer semi-mandatory/discretionary or discretionary activities. They had more trips and longer travel time overall. Students were also more likely to use active and public modes, and less likely to use private mode. Regarding active modes, the number of trips for students was less but they traveled longer distances. Non-workers were more likely to stay at home, and those who did travel tend to have a higher participation in activities overall; specifically, they participated in more semi-mandatory/discretionary and discretionary activities but fewer mandatory activities. While non-workers made more total trips than workers, they used active and public modes less. They made fewer trips, traveled shorter distances, and had shorter travel times by active modes. The only travel behavior that seemed to be elevated for this group was the number of trips by private modes.

Lastly, the neighborhood type of people's homes resulted in some significant differences in travel behaviors. Compared to urban residents, people living in more suburban or rural neighborhoods traveled longer distances and spent more time traveling overall. They used active and public modes less and private modes more. While they traveled longer distances and spent more time traveling by private mode, these trends were the opposite for active modes.

The following section describes and discusses results for air quality in more detail.

**Air Quality Results**

Table 4 and Table 5 present more complete results (coefficient estimates *B* and *p*-values) for the air quality measures in all of the models for urban and suburban/rural areas respectively: the 20 DVs, and the three air quality metrics (AQI number, AQI category, perceived air quality). Note that the two middle columns (AQI: yellow and orange) contain coefficients for the two AQI color categories (green as the base category) from the same set of models. In the following paragraphs, we interpret and discuss the air quality results for each type or set of activity/travel behavior outcomes.

**Table 4: Model results for air quality measures (urban)**

|  |  | *AQI* | | *AQI: Yellow* | | *AQI: Orange* | | *Perceived AQ* | |
|---|---|---|---|---|---|---|---|---|---|
| *Dependent variable* | *Model*[a] | *B* | *p* | *B* | *p* | *B* | *p* | *B* | *p* |
| Stayed at home (did not travel) | BL | -0.00115 | 0.815 | -0.2439 | 0.248 | 1.3821 | 0.104 | -0.1458 | 0.221 |
| Activities: Total out-of-home | QP | 0.00036 | 0.724 | 0.0376 | 0.374 | -0.1241 | 0.633 | -0.0053 | 0.814 |
|    Mandatory | QP | **0.00331** | **0.007** | **0.1601** | **0.002** | 0.0771 | 0.798 | 0.0256 | 0.349 |
|    Semi-mandatory/discretionary | QP | -0.00022 | 0.931 | -0.0592 | 0.572 | 0.0464 | 0.938 | -0.0090 | 0.872 |
|    Discretionary | QP | -0.00310 | 0.138 | -0.0534 | 0.529 | -0.5475 | 0.364 | -0.0640 | 0.168 |
| Number of trips (#): Total | QP | 0.00068 | 0.415 | 0.0404 | 0.238 | 0.0773 | 0.681 | -0.0015 | 0.932 |
| Distance traveled (miles): Total | LL | 0.00146 | 0.332 | 0.0216 | 0.726 | -0.2633 | 0.450 | -0.0368 | 0.266 |
| Travel time (minutes): Total | LL | **0.00260** | **0.016** | **0.0922** | **0.037** | 0.0815 | 0.744 | -0.0011 | 0.964 |
| Active mode user | BL | 0.00291 | 0.477 | 0.1226 | 0.470 | 0.4138 | 0.681 | -0.1416 | 0.129 |
|    Number of trips (#) | QP | 0.00087 | 0.574 | -0.0526 | 0.433 | *0.4158* | *0.058* | -0.0500 | 0.172 |
|    Distance traveled (miles) | LL | -0.00184 | 0.426 | -0.1272 | 0.190 | -0.1761 | 0.661 | **0.1289** | **0.016** |
|    Travel time (minutes) | LL | -0.00001 | 0.996 | -0.0907 | 0.401 | 0.5526 | 0.218 | -0.0693 | 0.246 |
| Public mode user | BL | -0.00386 | 0.498 | -0.1625 | 0.492 | 1.3158 | 0.172 | 0.0741 | 0.569 |
|    Number of trips (#) | QP | **0.00374** | **0.022** | 0.0938 | 0.195 | *0.4753* | *0.056* | 0.0516 | 0.222 |
|    Distance traveled (miles) | LL | 0.00119 | 0.626 | 0.0125 | 0.906 | 0.2590 | 0.516 | **0.1449** | **0.014** |
|    Travel time (minutes) | LL | **0.00874** | **0.003** | 0.2069 | 0.106 | 0.7188 | 0.135 | **0.1430** | **0.050** |
| Private mode user | BL | -0.00417 | 0.447 | 0.0792 | 0.731 | **-2.3259** | **0.025** | 0.0300 | 0.816 |
|    Number of trips (#) | QP | 0.00109 | 0.237 | 0.0505 | 0.175 | 0.1816 | 0.543 | 0.0143 | 0.474 |
|    Distance traveled (miles) | LL | *0.00271* | *0.097* | 0.0475 | 0.467 | -0.1368 | 0.808 | **-0.0709** | **0.044** |
|    Travel time (minutes) | LL | **0.00358** | **0.003** | **0.1202** | **0.014** | -0.0163 | 0.969 | 0.0169 | 0.520 |

[a] Models: BL = binary logit, QP = quasi-poisson, LL = log-linear
**Bold** if p<0.05; *italics* if p<0.10.

**Table 5: Model results for air quality measures (suburban/rural)**

|  |  | AQI | | AQI: Yellow | | AQI: Orange | | Perceived AQ | |
|---|---|---|---|---|---|---|---|---|---|
| Dependent variable | Model[a] | B | p | B | p | B | p | B | p |
| Stayed at home (did not travel) | BL | 0.00487 | 0.432 | 0.1995 | 0.439 | -11.7318 | 0.973 | 0.1746 | 0.221 |
| Activities: Total out-of-home | QP | 0.00028 | 0.806 | -0.0013 | 0.979 | -0.1725 | 0.574 | -0.0185 | 0.507 |
|     Mandatory | QP | 0.00039 | 0.798 | 0.0526 | 0.404 | -0.4436 | 0.293 | -0.0408 | 0.289 |
|     Semi-mandatory/discretionary | QP | -0.00096 | 0.698 | -0.0978 | 0.346 | 0.2032 | 0.759 | 0.0228 | 0.707 |
|     Discretionary | QP | 0.00127 | 0.570 | 0.0394 | 0.676 | -0.2040 | 0.735 | -0.0051 | 0.927 |
| Number of trips (#): Total | QP | 0.00074 | 0.428 | 0.0147 | 0.708 | -0.1085 | 0.660 | -0.0078 | 0.736 |
| Distance traveled (miles): Total | LL | **-0.00336** | **0.045** | -0.0645 | 0.358 | *-0.7055* | *0.077* | 0.0236 | 0.569 |
| Travel time (minutes): Total | LL | -0.00152 | 0.206 | -0.0123 | 0.807 | *-0.5154* | *0.071* | 0.0347 | 0.243 |
| Active mode user | BL | **0.01538** | **0.038** | **1.1431** | **0.001** | -10.9696 | 0.974 | 0.1642 | 0.392 |
|     Number of trips (#) | QP | 0.00409 | 0.342 | -0.0120 | 0.945 | NA | NA | 0.1595 | 0.100 |
|     Distance traveled (miles) | LL | 0.00444 | 0.443 | 0.2397 | 0.316 | NA | NA | 0.1419 | 0.267 |
|     Travel time (minutes) | LL | *0.01103* | *0.087* | 0.2745 | 0.303 | NA | NA | 0.2085 | 0.147 |
| Public mode user | BL | 0.00506 | 0.715 | 0.2755 | 0.610 | -10.0092 | 0.985 | 0.1884 | 0.550 |
|     Number of trips (#) | QP | 0.00080 | 0.858 | 0.0845 | 0.638 | NA | NA | -0.0358 | 0.715 |
|     Distance traveled (miles) | LL | 0.00811 | 0.246 | 0.1748 | 0.531 | NA | NA | *0.2666* | *0.071* |
|     Travel time (minutes) | LL | -0.00292 | 0.724 | -0.1606 | 0.632 | NA | NA | -0.0457 | 0.802 |
| Private mode user | BL | -0.02058 | 0.231 | -1.1664 | 0.166 | 9.4508 | 0.986 | -0.0991 | 0.828 |
|     Number of trips (#) | QP | 0.00047 | 0.617 | -0.0033 | 0.934 | -0.0570 | 0.816 | -0.0139 | 0.552 |
|     Distance traveled (miles) | LL | *-0.00320* | *0.058* | -0.0576 | 0.416 | *-0.7111* | *0.076* | 0.0181 | 0.665 |
|     Travel time (minutes) | LL | -0.00179 | 0.156 | -0.0281 | 0.594 | -0.4728 | 0.114 | 0.0295 | 0.345 |

[a] Models: BL = binary logit, QP = quasi-poisson, LL = log-linear
**Bold** if p<0.05; *italics* if p<0.10. NA if the coefficient was unable to be estimated due to the small sample size.

To begin, there were no significant associations between air quality and whether or not someone stayed at home or traveled for both urban and suburban/rural areas. Although not significantly different from zero ($p = 0.10$), the magnitude of the estimated coefficient for orange AQI among urban residents was fairly large, suggesting that people who experienced an orange air quality day were almost four times as likely to stay at home (odds ratio $OR = e^B = 3.98$) than on a day with green air quality. Recall that this situation reflects only 1% of the person-day observations in the dataset, so the study may have lacked the power to detect a significant effect for this (and other) outcomes on orange days.

There were few consistent patterns of association found between measures of air quality and activity participation. In both urban and suburban/rural areas, total activities, as well as semi-mandatory/discretionary activities, both did not seem to be linked to air quality. The only significant associations were for mandatory activities (work and school) in urban areas: the models showed positive associations with AQI number and yellow AQI, implying that people tended to participate in slightly more mandatory activities on days with more air pollution or on yellow (vs. green) air quality days. Specifically, in urban areas, the model predicts a 3% increase in mandatory activities for every 10-point increase in AQI ($e^{10B} = 1.03$), and 17% greater participation in mandatory activities on yellow days compared to green days ($e^B = 1.17$). We are unsure how to explain this finding. In northern Utah, air pollution levels often start to elevate during clear days after a snowstorm, so it could be that some workers or students were not commuting on snowy days and started to on clear days when the air quality turned to yellow. While not significant, it is notable that the estimated coefficients among urban residents for discretionary activities were negative in all of the models. If true, this could imply that, on days with elevated levels of perceived or measured air pollution, people tend to forego discretionary activities like shopping, eating out, or indoor/outdoor exercise. This would match our expectation that the need for and scheduling of discretionary activities is more flexible; they could be shifted to other (better air quality) days or even canceled. Urban residents have more flexibility due to greater accessibility.

The air quality metrics showed links with a few total (all-mode) travel behavior outcomes. In urban areas, the only (marginally) significant association was between AQI and yellow AQI days with travel time. Specifically, the model predicts a 10% increase in total travel time ($e^B = 1.10$), on average, when comparing yellow to green air quality days, or a 3% increase ($e^{10B} = 1.03$) for every 10-point increase in AQI. In suburban/rural areas, there were significant associations between AQI number and orange AQI days for distance traveled and travel time (marginally significant). The model predicts a 3% reduction in total miles traveled for every 10-point increase in AQI ($e^{10B} = 0.97$), and a 50% reduction in total miles traveled ($e^{10B} = 0.49$) on orange days. Meanwhile, there was a marginally significant association between orange AQI days and travel time in suburban/rural areas. The model results predict a 40% decrease in total travel time for orange AQI days ($e^{10B} = 0.60$). These are substantial declines that (as seen later) appear to be being driven by large decreases in travel amounts for private (automobile) modes among suburban/rural residents. If true, this would be quite promising evidence for efforts to reduce emissions from polluting modes on days with poor air quality. However, recall that the sample size for this orange situation is quite low: only 23 person-days.

Turning to active modes of transportation (walking and bicycling), some results were significant. In urban areas, the models present some evidence that the use of active modes increased on days with worse AQI. Among active travelers, the model showed large-magnitude increases in travel on orange air quality days: 51% more trips ($e^B = 1.52$). The model also shows that distance traveled by active modes increased on days with poorer perceived air quality. Specifically, urban residents walked or bicycled 14% more ($e^B = 1.14$) on days with one-point worse air quality (on a five-point scale). For residents of suburban/rural areas, the models' results also show some evidence that the use of active modes increased on days with worse air pollution. Specifically, these respondents had 17% greater odds ($OR = 1.17$) of using active modes for every 10-point increase in AQI, or more than three times as likely ($OR = 3.14$) on yellow as compared to green air quality days. The models' results also show an average 12% increase ($e^{10B} = 1.12$) in travel time spent walking or biking for a 10-point increase in AQI. There were also positive (albeit not statistically significant) associations between active mode use and perceived air pollution among suburban/rural residents. Overall, these results support an altruistic response to air pollution, although the magnitudes of the effects on orange days should be viewed with caution. Also, the results suggest different responses by neighborhood type: Suburban/rural residents were more likely to be active mode users, while urban residents were more likely to increase their use of active modes.

Model results for public transit modes (bus only) in urban areas imply similar altruistic responses to air pollution. No coefficients were significant for choosing to ride the bus, but there were several significant associations between AQI categories (AQI number, orange AQI, and perceived AQI) and travel behavior outcomes (number of trips, distance traveled, and travel time). The model results show that the number of trips by public mode increased on days with poor air quality: a 4% increase for every 10-point increase in AQI ($e^{10B} = 1.04$), and a 61% increase on orange (versus green) air quality days ($e^B = 1.61$). There was also a 9% increase in travel time by bus for every 10-point increase in AQI ($e^{10B} = 1.09$). Additionally, the model results show that urban residents spent more time and longer distance traveling using public modes on days greater perceived air pollution: The model predicts a 15–16% increase in distance traveled and travel time. In suburban/rural areas, however, we did not find any significant results for public mode, except for a marginally-significant association between perceived AQI and distance traveled. Based on the results, respondents rating the air quality 1-point worse might be expected to increase their transit distance traveled use by 31% ($e^B = 1.31$). Altogether, there is some evidence that transit riders in our sample tended to use the bus more on days with worse (measured or perceived) air quality. However, this evidence was concentrated among urban residents, suggesting that transit access and availability might be pre-conditions for being able to change travel behaviors.

Finally, we come to private mode (automobile) use. In urban areas, there were significant associations between all AQI categories (AQI number, yellow AQI, orange AQI, and perceived AQI) and most travel behavior outcomes (mode users, distance traveled, and travel time). We can see that distance traveled and travel time by private modes increased on days with poor air quality. The results show a 3% increase in the distance traveled ($e^{10B} = 1.03$), and a 4% increase in travel time ($e^{10B} = 1.04$), for every 10-point increase in AQI. Also, on yellow (versus green) days, the model predicts a 13% increase in travel time by private modes. Even air quality perception had a meaningful and significant impact; people rating air quality 1-point worse might be expected to decrease the distance traveled by private modes by 7% ($e^B = 0.93$). On orange days, for urban residents, the model showed a large and statistically-significant 90% decrease in the odds ($e^{10B} = 0.10$) of someone being a private mode user; however, recall the sample size limitation. In suburban/rural areas, we only saw two marginally significant associations: between the AQI number and orange AQI days with distance traveled. The model predicted that private mode users drove 3% fewer miles for every 10-point increase in AQI ($e^{10B} = 0.97$), and on orange days a decrease of 51% ($e^{10B} = 0.49$). Recall the small sample size, but also remember the marginally-significant decrease in total (all-mode) distance traveled by suburban/rural residents on these days, mentioned earlier. Changes in driving amounts seems to be affecting this result.

**CONCLUSION**

In this study, we sought to determine whether and how measured or perceived levels of air pollution affected individuals' daily activity participation and travel behaviors in urban and suburban/rural areas. Although there were not many significant air quality coefficients estimated by the models of activity and travel outcomes (Table 4), there were enough for us to make some overarching conclusions. Activity and travel behavior response patterns in both urban and suburban/rural areas exhibited both similarities and differences.

First, the only change in activity participation that we observed was that participation in mandatory activities (work and school) appeared to increase on days with worse air pollution in urban areas. While we speculated about this potentially being a side-effect of shifting work or school travel to different days, we are unsure of this result and encourage additional research to identify a more convincing explanation. There was some but not convincing evidence that urban residents made fewer discretionary trips on poor air quality days, which, if true, could imply that greater multimodal accessibility could allow greater flexibility in activity schedules.

Second, there appeared to be some detectable changes in traveler behaviors on days with poorer measured air quality, but the effects were different for active, public, and private modes, and for residents of urban and suburban/rural neighborhoods. For active modes in urban areas, not much changed. Meanwhile, people living in suburban/rural neighborhoods were more likely to use active modes (and perhaps increase their active travel duration) as air pollution increased. In contrast, air pollution did not appear to encourage more people to shift to using public transit, but existing transit users in urban areas tended to ride the bus more frequently and for a longer time, whereas no changes in public mode use behavior were measured for residents of suburban/rural areas, likely due to the greater difficulty accessing transit services. For private mode use, in urban areas, people appeared to spend more time driving on poor air quality days, whereas in suburban/rural areas, there was some evidence of fewer miles driven.

Third, there were some very large (and sometimes significant) measured changes in travel behaviors on unhealthy (orange) air quality days: more walking/bicycling (urban only), more transit use and users (urban only), and less driving (all areas, especially suburban/rural) and fewer automobile users (urban only). However, the small sample size calls into question the validity of the estimates. Despite our best efforts, our study's natural experiment suffered from a weak "treatment effect." Because few people were exposed to an orange air quality day, our study likely lacked sufficient power to detect significant effects of days with unhealthy air pollution. For instance, we were unable to estimate coefficients for active and public mode travel behaviors in suburban/rural areas on orange days due to sample size limitations.

Fourth, people who perceived the air quality to be worse tended to use active and public transit modes more (longer distances and travel times), but this is only among users of these modes and mostly among urban residents. There were no results suggesting that air quality perceptions shifted people towards walking, bicycling, or riding the bus. It could be that both measured (and announced) air pollution and perceptions of air quality affect peoples' behavioral responses in slightly different ways. We encourage more research investigating air quality perceptions.

Overall, these results are somewhat encouraging for behavioral responses to air pollution. There is more evidence of altruistic responses than risk-averse responses: more people choosing active modes, more bus use among transit riders, not more people using automobiles, and potentially dramatic shifts on days with much worse air pollution. This finding suggests that policies to spread awareness of the harms of air pollution from automobile emissions and other "soft" travel behavior change strategies might be able to encourage people to choose less polluting modes on poor air quality days.

Notably, we observed that, in urban areas, active and public transportation modes were used more frequently on polluted days, possibly due to the closer proximity of destinations and transit accessibility, making these more feasible travel options. Conversely, in suburban/rural areas, we did not witness a considerable change (except for a greater chance of using active modes), which could be largely attributable to the built environment with larger distances between developed areas, rendering them less walkable for daily travel to work, school, shopping, etc. Furthermore, we should acknowledge that the non-significant changes observed in suburban/rural areas may partly result from the smaller size of these populations in our sample.

We recommend several efforts for future research to advance upon this study. First, most of our participants did not experience a day of "unhealthy" air quality (orange or worse), so it was difficult to detect significant behavioral shifts due to air pollution. Future studies should try to capture a wider range of air quality levels, although this is difficult when relying upon unpredictable atmospheric conditions. Second, our use of self-report travel diaries had a high respondent burden, was potentially prone to reporting errors, required much data cleaning, and limited the number of days we could study. The use of a GPS-based travel survey could help mitigate many of these issues, and it might also allow for a longer study period to hopefully capture more variation in air pollution levels. Third, our analysis itself could be improved through more advanced statistical methods. For example, we did not account for various natures of our dataset: panel (people observed over multiple days), multilevel (people within households within neighborhoods), or multivariate (multiple potentially-correlated dependent variables). In future work, advances such as these could help provide stronger evidence of how activity and travel behaviors are affected by episodes of area-wide air pollution.

## ACKNOWLEDGEMENT

The work presented in this paper conducted with support from Utah State University and the Mountain-Plains Consortium, a University Transportation Center funded by the U.S. Department of Transportation. The study was reviewed and approved by the Utah State University Institutional Review Board (Protocol #9246). The contents of this paper reflect the views of the authors, who are responsible for the facts and accuracy of the information presented. Special thanks to David Christensen, Peter Gilbert, Adam Pack, and Joshua Ward for helping to review literature and prepare the survey questions; Seth Thompson and Joshua Ward for administering the survey; Joshua Ward for helping to clean the data; and Joshua Ward and Prasanna Humagain for conducting a preliminary version of this analysis.

## AUTHOR CONTRIBUTIONS STATEMENT

The authors confirm contribution to the paper as follows: study conception and design: PAS, FSM; data collection: PAS; analysis and interpretation of results: FSM, MVS, PAS; draft manuscript preparation: FSM, MVS, PAS. All authors reviewed the results and approved the final version of the manuscript.

9. Holmes, A. M., Lindsey, G., & Qiu, C. (2009). Ambient air conditions and variation in urban trail use. *Journal of Urban Health, 86*(6), 839-849. https://doi.org/10.1007/s11524-009-9398-8
10. Acharya, S., & Singleton, P. A. (2022). Associations of inclement weather and poor air quality with non-motorized trail volumes. *Transportation Research Part D: Transport and Environment, 109*, 103337. https://doi.org/10.1016/j.trd.2022.103337
11. Henry, G. T., & Gordon, C. S. (2003). Driving less for better air: Impacts of a public information campaign. *Journal of Policy Analysis and Management, 22*(1), 45–63. https://doi.org/10.1002/pam.10095
12. Wang, S. Y., Hipps, L. E., Chung, O. Y., Gillies, R. R., & Martin, R. (2015). Long-term winter inversion properties in a mountain valley of the western United States and implications on air quality. *Journal of Applied Meteorology and Climatology, 54*(12), 2339-2352. https://doi.org/10.1175/JAMC-D-15-0172.1
13. Utah Department of Environmental Quality (Utah DEQ). (n.d.) *Air quality forecast: Forecast legend*. https://air.utah.gov/forecastLegendAQI.html#Action
14. Utah Department of Transportation (UDOT). (n.d.). *TravelWise: Rethink your trip*. https://travelwise.utah.gov/
15. Environmental Protection Agency. (2018). *Smart Location Mapping*. https://www.epa.gov/smartgrowth/smart-location-mapping
16. Humagain, P., & Singleton, P. A. (2021). Impacts of episodic poor air quality on trip-making behavior and air quality perceptions from a longitudinal travel diary survey. Presented at the 100th Annual Meeting of the Transportation Research Board, Washington, DC.
17. AirNow. (n.d.). Air quality index (AQI) basics. https://www.airnow.gov/aqi/aqi-basics/
18